\documentclass[apjl]{emulateapj}

\begin{document}

\title{AN OPTICAL--NEAR--INFRARED OUTBURST WITH NO ACCOMPANYING $\gamma$-RAYS IN THE BLAZAR PKS 0208--512}

\author{RITABAN CHATTERJEE\altaffilmark{1,5}, G. FOSSATI\altaffilmark{2}, C. M. URRY\altaffilmark{3}, C. D. BAILYN\altaffilmark{1}, L. MARASCHI\altaffilmark{4}, M. BUXTON\altaffilmark{1}, E. W. BONNING\altaffilmark{3,6},  J. ISLER\altaffilmark{1}, P. COPPI\altaffilmark{1}}

\altaffiltext{1}{Department of Astronomy, Yale University, PO Box 208101, New Haven, CT 06520-8101, USA.}
\altaffiltext{2}{Department of Physics and Astronomy, Rice University, 6100 Main St., Houston, TX 77005, USA.}
\altaffiltext{3}{Department of Physics and Yale Center for Astronomy and Astrophysics, Yale University, PO Box 208121, New Haven, CT 06520-8121, USA.}
\altaffiltext{4}{INAF--Osservatorio Astronomico di Brera, V. Brera 28, I-20100 Milano, Italy.}
\altaffiltext{5}{Current address: Department of Physics and Astronomy 3905, University of Wyoming, 1000 East University, Laramie, WY 82071, USA; rchatter@uwyo.edu}
\altaffiltext{6}{Current address: Quest University Canada, 3200 University Boulevard Squamish, BC V8B 0N8, Canada.}

\begin{abstract}
We report the discovery of an anomalous flare in a bright blazar, namely, PKS 0208--512, one of the targets of the Yale/SMARTS optical--near IR (OIR) monitoring program of \textit{Fermi} blazars. We identify three intervals during which PKS 0208--512 undergoes outbursts at OIR wavelengths lasting for $\gtrsim$3 months. Its brightness increases and then decreases again by at least 1 magnitude in these intervals. In contrast, the source undergoes bright phases in GeV energies lasting $\gtrsim$1 month during intervals 1 and 3 only. The optical--near IR outburst during interval 2 is comparable in brightness and temporal extent to the OIR flares during intervals 1 and 3 which do have $\gamma$-ray counterparts. By analyzing the $\gamma$-ray, optical--near IR, and supporting multi-wavelength variability data in details, we speculate that the OIR outburst during interval 2 was caused by a change in the magnetic field without any change in the total number of emitting electrons or Doppler factor of the emitting region. Alternatively, it is possible that the location of the outburst in the jet during interval 2 was closer to the black hole where the jet is more compact and the bulk Lorentz factor of the material in the jet is smaller. We also discuss the complex optical--near IR spectral behavior during these three intervals.
\end{abstract}

\keywords{black hole physics --- galaxies: active --- galaxies: individual (PKS 0208--512) ---  galaxies:jets --- quasars: general --- radiation mechanisms: non-thermal}

\section{Introduction}
Jets of blazars are often bright over a broad wavelength range (from radio to $\gamma$-rays in extreme cases) and highly time-variable. Hence, multi-wavelength monitoring is an important technique to understand the physics of jets. Data from the \textit{Fermi Gamma-Ray Space Telescope} and supporting multi-wavelength monitoring programs at other wavebands have generated important recent studies of blazar jets \citep{abd10_SED}. The bright source list from the \textit{Fermi} 2-yr catalog (2FGL) contains more than 1000 active
galactic nuclei, most of which are blazars \citep{nol12,ack_2yrAGN}. The number of sources is large and the variability timescale in many flat spectrum radio quasars (FSRQs) is similar at $\gamma$-ray and optical--near IR (OIR) bands. Therefore, one should ideally design OIR monitoring programs such that the sampling frequency of a source at OIR and $\gamma$-ray energies are similar while maintaining a regular cadence for many blazars.

Here we report the discovery of an anomalous flare in a bright blazar, one of the targets of one such project, namely, the Yale/SMARTS optical--near IR monitoring program\footnote{http://www.astro.yale.edu/smarts/fermi/} \citep{bon12,cha12}. Among other \textit{Fermi}-LAT monitored blazars in the southern sky, we have followed the variations in emission of PKS 0208--512, a blazar at redshift z=1.003 \citep{hea08}. It was discovered in the Parkes Survey of Radio Sources \citep{bol64} and was detected regularly \citep{tho95} by EGRET on board the \textit{Compton Gamma Ray Observatory (CGRO)}. Most recently, multi-wavelength observations of its large-scale jet were presented in \citet{per11}. PKS 0208--512 was originally classified as a BL Lac object based on the equivalent width of MgII line \citep{hea08}. Recently, \citet{ghi11} pointed out that its spectral energy distribution (SED) resembles that of FSRQs and its broad lines are very luminous but are overwhelmed by the brighter continuum luminosity \citep[$L_{MgII}\sim$10$^{44}$ erg s$^{-1}$;][]{sca97}. Hence, it should be classified as an FSRQ. 
\begin{figure*}
\epsscale{.70}
\plotone{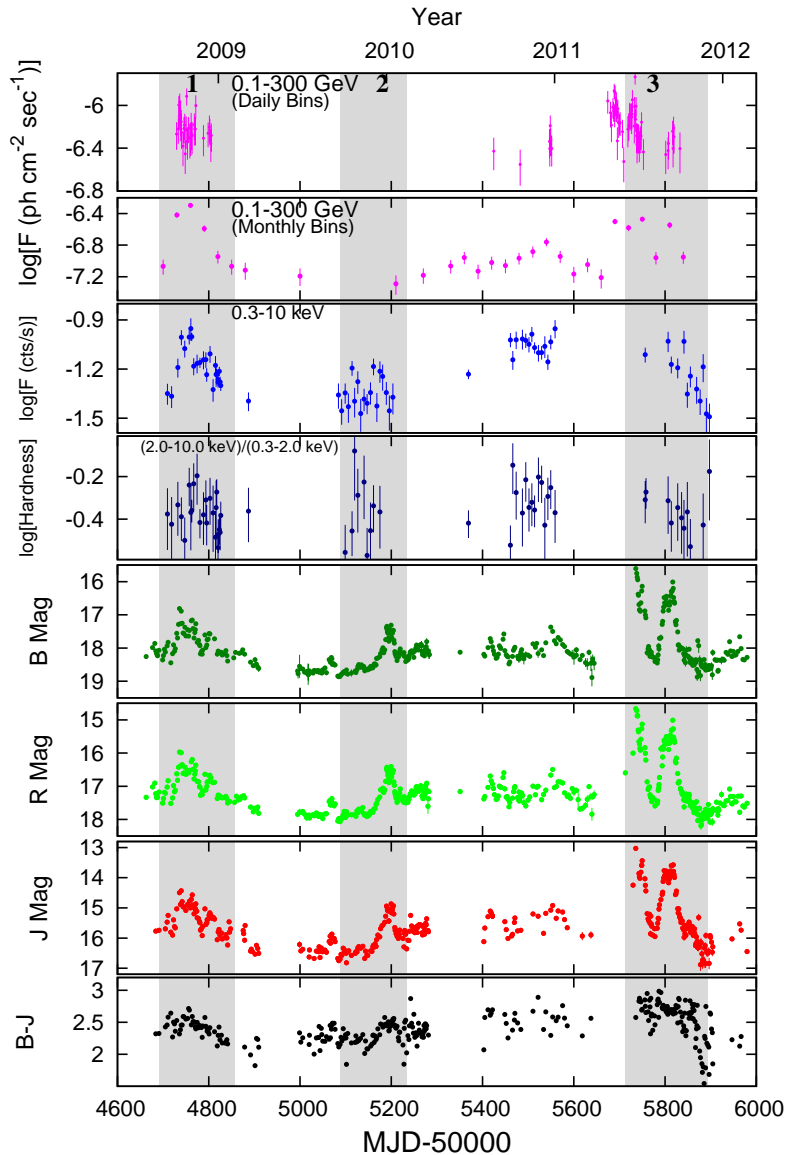}
\caption{Variation of 0.1-300 GeV $\gamma$-ray flux from \textit{Fermi}-LAT in daily and monthly bins, X-ray flux and hardness ratio from \textit{Swift}-XRT, $B-$, $R-$ and $J-$band flux densities, and $B-J$ color (from ANDICAM on SMARTS 1.3 m telescope) of PKS 0208--512 from 2008 August to 2012 February. The gray shaded regions indicate three intervals during which the source underwent long-term and powerful outbursts at optical--near IR wavelengths. While the GeV flux increased during intervals 1 and 3, the source was detected at a very low $\gamma$-ray flux level at only one monthly bin during the second interval.}
\label{lc_all}
\end{figure*}

PKS 0208--512 was detected by the Large Area Telescope (LAT) on board \textit{Fermi} at a level higher than $5 \times$10$^{-7}$ ph cm$^{-2}$ sec$^{-1}$ with a time-binning of 1 day regularly from 2008 September to 2008 December and from 2011 April to 2011 September. It was too close to the Sun at the start of the latter brightening phase. We obtained well-sampled optical and near-infrared light curves starting 2011 June, when it became observable from CTIO, Chile as well as the entirety of the interval from 2008 September to 2008 December. In addition, we monitored this blazar with one observation per night during another OIR bright phase from 2009 November to 2010 January during which it was quiescent in $\gamma$-rays. This OIR outburst is comparable in brightness and temporal extent to the OIR flares in the intervals mentioned above during 2008 and 2011 which do have $\gamma$-ray counterparts. In this Letter, we analyze the $\gamma$-ray and optical light curves to investigate, in particular, the physical mechanism that can produce this optical--near IR-only outburst.

In Section 2, we present the observations and data reduction procedures. In Section 3, we describe the results. We discuss possible scenarios that can explain the nature of the observed $\gamma$-ray/OIR variability in Section 4.\\ \\

\section{Data}
\subsection{GeV Data}
We derived the 0.1--300 GeV $\gamma$-ray flux of PKS 0208--512 by analyzing data from \textit{Fermi-LAT} using the standard \textit{Fermi-LAT Science Tools} software package (version v9r27p1). We analyzed a Region of Interest of 15$^\circ$ in radius, centered at the position of PKS 0208--512, using the maximum likelihood algorithm implemented in \textit{gtlike}, modeling sources with a simple power law. We included all sources within 15$^\circ$ of PKS 0208--512, extracted from the \textit{Fermi} 2-yr catalog (2FGL), with their normalizations kept free and spectral indices fixed to their catalog values. We use the currently recommended P7SOURCE\_V6 set of the instrument response functions, Galactic diffuse background model, and isotropic background model. In the \textit{Fermi} LAT Second Source Catalog, PKS 0208--512 has been modeled with a simple power-law of index 2.39$\pm$0.04 and its curve significance is 1.6. This value of the curve significance is much lower than 16.0 which indicates switching to a curved spectrum such as log parabola will not improve the fit significantly. Hence, we fix the spectral index of PKS 0208--512 at 2.39 in the model file used in the \textit{gtlike} analysis. 
\begin{figure*}
\epsscale{0.7}
\plotone{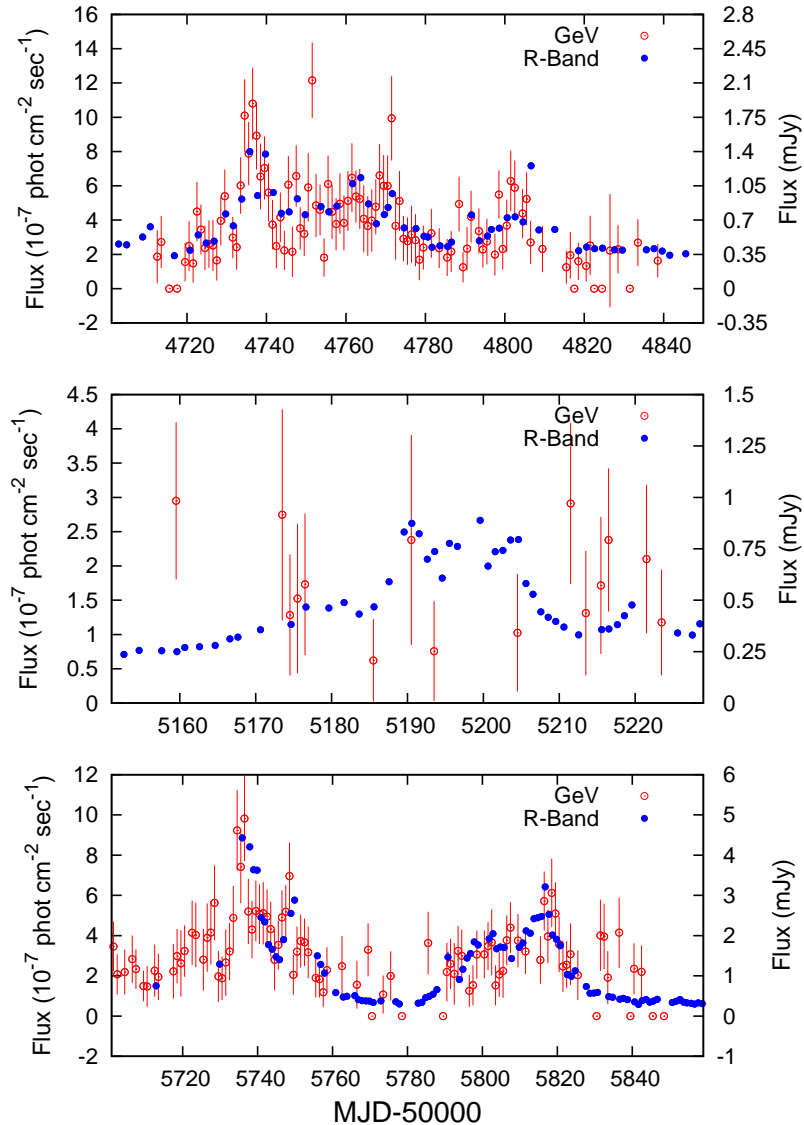}
\caption{0.1-300 GeV $\gamma$-ray flux (red open circles) and optical $R-$band flux density (blue filled circles) of PKS 0208--512 during intervals 1 (top), 2 (middle) and 3 (bottom). These GeV light curve segments are binned over 1-day intervals and include flux values for detections with TS $>$ 4.0 (equivalent to $\sim$2$\sigma$). The left and right hand vertical axes denote the units of the GeV and R-band flux, respectively. It is evident that the GeV and $R-$band variability are remarkably well-correlated during intervals 1 and 3.}
\label{lc_zoom13}
\end{figure*}

The detection significance of PKS 0208--512 in 2FGL, which integrates the first 24 months of data, is 35.8$\sigma$. We calculate the fluxes of PKS 0208--512 from 1-day integrations during 2008 September to 2012 February, with a detection criterion that the maximum-likelihood test statistic (TS) exceed 25.0. A TS value of 25.0 is roughly comparable to a 5$\sigma$ detection. This light curve is shown in the top panel of Figure \ref{lc_all}. The GeV light curve is more sparsely sampled than that in the optical--near IR frequencies. This is because the blazar was not detectable at TS $>$ 25.0 level with 1-day integrations for long periods during the 3.5 yr time interval considered here. To investigate the nature of GeV variability of this blazar during intervals when it was not detected in 1-day bins, we also calculate the GeV fluxes from 30-day integrations during 2008 September to 2012 February. We show the monthly light curve in the second panel of Figure \ref{lc_all}. The gaps in the monthly light curve consist of bins when PKS 0208--512 was not detected at TS $>$ 25.0 level even with a 30-day integration.
\begin{figure*}
\epsscale{1.0}
\plotone{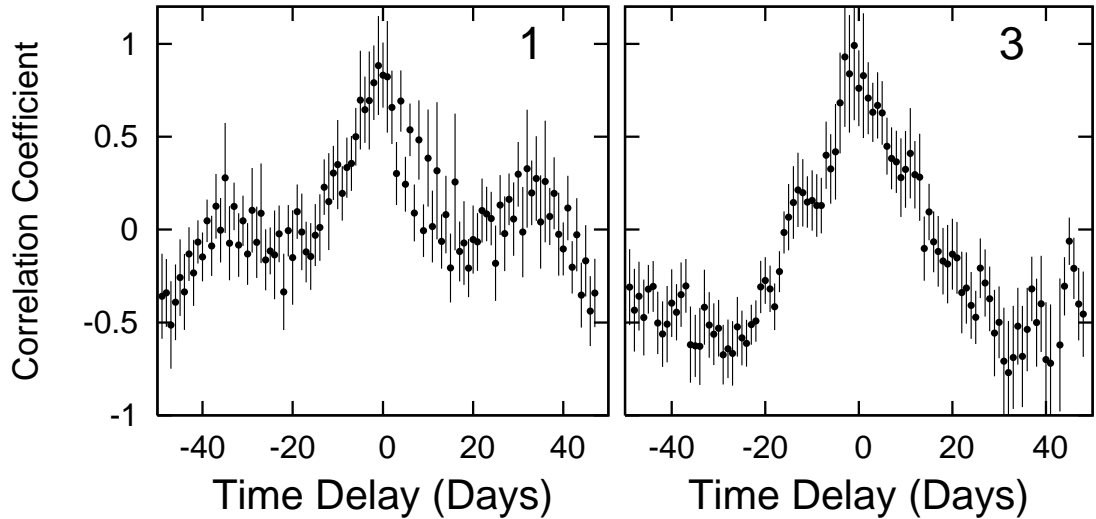}
\caption{Discrete cross-correlation function (DCCF) of the $\gamma$-ray and $R$-band light curves of PKS 0208--512 during intervals 1 and 3. The time delay is defined as positive if the $R-$band variations lead those at $\gamma$-ray energies. During both intervals, the $\gamma$-ray and $R-$band variations are very strongly correlated with zero time lag.}
\label{dccf_zoom13}
\end{figure*}

We identify three intervals during which PKS 0208--512 undergoes outbursts at OIR wavelengths lasting for $\gtrsim$3 months with date ranges i) MJD 54690--54856 (2008 August to 2009 January) ii) MJD 55087--55233 (2009 September to 2010 February) and iii) MJD 55712--55893 (2011 May to 2011 November). These intervals are shown as the gray shaded regions in Figure \ref{lc_all}. These intervals are defined such that i) they contain a steady rise of flux by 1.3 magnitudes or more ii) they contain the corresponding decaying branch down to the ``quiescent" level at which the rise started iii) in the cases where another steady rise of $\sim$ 0.5 magnitude or larger started before the flux level decreased to the ``quiescent" level, then the interval is cut off at the start of the said rise and iv) the length of each interval is 2 months or more. While defining ``steady" rise or decay, we ignore small fluctuations in the rising or decaying branch which are less than 0.5 magnitudes and/or less than a month long because our goal in this work is to investigate the intervals containing longer-term and more powerful outbursts. We include two large OIR outbursts together in interval 3 because they were very close in time and only the decaying branch of the first flare was fully sampled in OIR frequencies. We assume that for our purpose counting these two subsequent outbursts into one interval simplifies the interval definitions without any loss of information. We test this in subsequent analysis. 

In contrast to the three intervals containing large OIR outbursts, the source undergoes bright phases in GeV energies lasting $\gtrsim$1 month only during intervals 1 and 3. While during intervals 1 and 3, the blazar was detected at numerous daily bins and all monthly bins, it was detected in no daily bin and only one monthly bin during interval 2. The average GeV flux during the monthly bin was ($5.1\pm1.4) \times$10$^{-8}$ ph cm$^{-2}$ sec$^{-1}$. This is more than an order of magnitude fainter than that in intervals 1 and 3. To better compare optical and $\gamma$-ray variations, we also obtained a GeV light curve of this blazar with a detection criterion of TS $>$ 4.0 (comparable to a 2$\sigma$ detection). We plot this light curve for all three intervals along with $R-$band variability in Figure \ref{lc_zoom13}.

\subsection{Optical--Near--IR and X-Ray Data}
All of the measurements in $B-$, $R-$, and $J-$band are from the ANDICAM instrument on SMARTS 1.3m telescope located at CTIO, Chile. ANDICAM is a dual-channel imager with a dichroic that feeds an optical CCD and an IR imager, which can obtain simultaneous data at one optical and one near-IR band. For details of data acquisition, calibration and data reduction procedures, see \citet{bon12}. 
\begin{figure*}
\epsscale{0.7}
\plotone{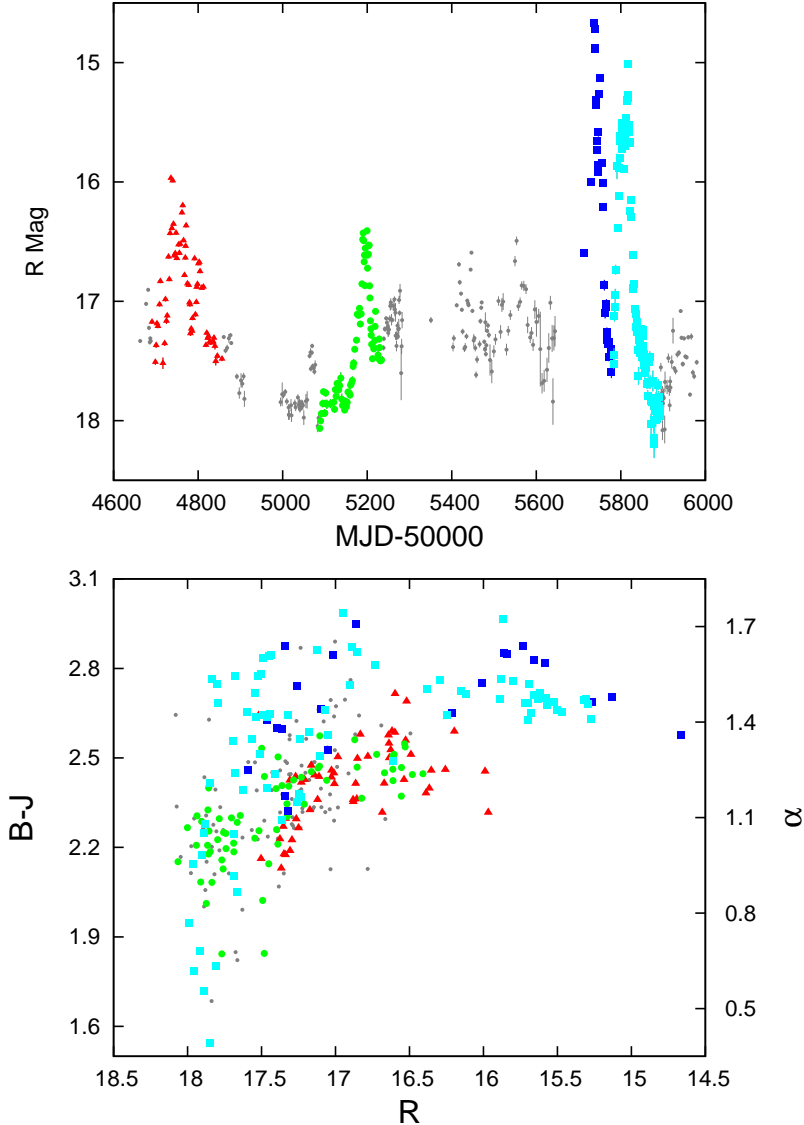}
\caption{Top panel shows the $R-$band light curve of PKS 0208--512 with intervals 1, 2, and 3 denoted by red triangles, green circles, and blue or cyan squares, respectively. Data points outside of the three intervals are denoted by smaller gray filled circles. The bottom panel shows the $B-J$ vs. $R$ color-magnitude diagram in which the right hand vertical axis denotes the corresponding $B-J$ spectral index. Color codes are the same as the top panel. Color variability looks remarkably similar during intervals 1 and 2, while the blazar is much redder and brighter during interval 3.}
\label{colormag}
\end{figure*}

We obtained the X-ray flux and hardness ratio during the same interval from the \textit{Swift}-XRT monitoring program of \textit{Fermi}-LAT sources of interest\footnote{http://www.swift.psu.edu/monitoring/}. Most of the observations were carried out with exposure times of $\sim$1 ks. The mean count rate was 0.07 cts/s in 0.3-10.0 keV energy range and the mean background count rate was 0.0007 cts/s during this entire period. As in the optical, X-ray variability was present during all three intervals. We use the \textit{Swift}-XRT data products generator\footnote{http://www.swift.ac.uk/user\_objects/docs.php} \citep{eva09} to derive the average photon index during the three relevant intervals. The photon index stays between 1.58 and 1.75, indicating no significant difference between intervals. The gaps in the X-ray light curve denotes intervals during which PKS 0208--512 was not observed by \textit{Swift}. $B-$, $R-$, and $J-$band light curves, and the variation of the $B-J$ color, X-ray flux and hardness ratio are shown in Figure \ref{lc_all}.

\section{Results}
\subsection{$\gamma$-Ray/Optical Correlation in Intervals 1 and 3}
From Figure \ref{lc_zoom13}, it is evident that the GeV and $R-$band variability are remarkably well-correlated during intervals 1 and 3. To confirm this, we cross-correlate the $\gamma$-ray and $R-$band light curves using the discrete cross-correlation function \citep[DCCF;][]{ede88}. A similar cross-correlation analysis for interval 2 was not possible due to infrequent detection even at TS $>$ 4.0 level. As shown in Figure~\ref{dccf_zoom13}, the $\gamma$-ray and $R$-band variability in this blazar during both intervals are very strongly correlated with zero time lag. This is consistent with similar correlated $\gamma$-ray and optical/near-infrared variability seen in many other blazars \citep[e.g.,][]{bon09,bon12}. The degree of correlation and lack of a significant lag can be explained by the standard leptonic scenario in which the $\gamma$-rays and OIR emission are generated by the same relativistic electrons in the jet through inverse-Compton and synchrotron processes, respectively. The source of the seed photons that are being scattered may be the synchrotron photons generated within the jet, in which case it is termed synchrotron self-Compton (SSC) process \citep{mar92,chi02,arb05}, or from outside the jet (radiation from broad emission line region, accretion disk, or dusty torus), termed external Compton or EC process \citep{sik94,cop99,bla00,der09}. Both SSC and EC scenarios predict strong correlation with zero time lag.

\subsection{Absence of Correlation in Interval 2}
As shown in Figure \ref{lc_all}, the source underwent an optical--near IR outburst during interval 2. The increase of brightness was comparable to that in interval 1 and the $B-J$ color variations (bottom panel of Figure \ref{lc_all}) were similar to those in intervals 1 and 3. But unlike those flares, in interval 2, there was no significant increase in the GeV flux and the average GeV flux was smaller than average flux values of intervals 1 and 3 by at least an order of magnitude.

To try to understand the physical differences among these flares, we investigated more closely the OIR spectral index and color in intervals 1, 2, and 3. Figure \ref{colormag} shows the $B-J$ vs. $R$ color magnitude diagram as well as the corresponding $B-J$ spectral index along with the $R-$band light curve. Surprisingly, there is no clear difference in the above properties between intervals 2 and 1. \citet{bon12} showed that many blazars show redder $B-J$ colors when brighter and suggested this was because the jet is redder than the accretion disk. When the jet gets brighter -- and the jet has much larger amplitude variations than the disk -- the overall color reddens. PKS 0208--512 shows this familiar trend in the color-magnitude diagram during flares 1 and 2. In flare 3, the color reddens as the source brightens, but it remains much redder than in flares 1 and 2. At the brightest end, the color changes are minimal, again unlike flares 1 and 2. Comparable trends are seen in many blazars \citep{bon12}. This may occur when the jet becomes so much brighter than the disk that a further increase in jet brightness does not affect the color anymore.

It is possible that during interval 3, the synchrotron peak of the SED is at a lower frequency than it was during intervals 1 and 2. This would make the near-IR portion of the SED steeper. Hence, the bluer disk emission will be more dominant in the OIR bands when the jet emission is weak. This may cause $B-J$ values to be redder during the brighter phase and bluer during the fainter phase than intervals 1 and 2 as observed. 

Note that there is no significant difference in the nature of the GeV/$R-$band correlation in the two consecutive flares during interval 3 (Figure \ref{lc_zoom13}). In addition, the data points corresponding to those two flares overlap with each other (Figure \ref{colormag}) indicating no significant difference in the color-magnitude behavior. This is consistent with including them together as part of the same interval.

\section{Discussion}
The behavior of PKS 0208--512 shows complex spectral behavior during three strong optical/near-infrared flares. We considered what physical process(es) could explain the anomalously low $\gamma$-ray flux during interval 2, even as the OIR colors remain very similar.

There is one scenario in which a change at optical--near IR wavelengths is not accompanied by a change in the GeV energies, namely, the optical synchrotron emission is due to a change in the magnetic field only, i.e., its magnitude and/or direction, while the GeV emission is due to a temporarily steady external Compton process. Inverse-Compton emission depends on the total number of emitting electrons (N${\rm _e}$), Doppler factor ($\delta$), and the number of seed photons available for scattering while synchrotron emission depends on N${\rm _e}$, $\delta$, as well as the magnetic field and the viewing angle in the form Bsin$\phi$, where viewing angle ($\phi$) is defined as the angle between the direction of the magnetic field vector and the line of sight. Variability in the observed $\gamma$-ray and OIR emission can be generated by changes in any one or a combination of these parameters. Hence, we conclude that the OIR outburst during interval 2 might be caused by a change in the magnetic field in the emitting region without any change in the other two parameters described above. 

The critical frequency ($\nu_{\rm peak}$) at which most of the synchrotron emission occurs is proportional to the magnetic field (Bsin$\phi$) while the synchrotron flux is proportional to (Bsin$\phi$)$^2$. The $R-$band flux increases by $\sim$1.4 mag during interval 2 which is equivalent to an increase of brightness by a factor of $\sim$3.6. The required change in Bsin$\phi$ for this to be caused by a change in magnetic field only is an increase by a factor $\sim$1.9. While this will increase $\nu_{\rm peak}$ by the same factor, its observational detection is difficult due to i) the uncertainty in the exact value of $\nu_{\rm peak}$ in FSRQs including PKS 0208--512, and ii) the possible presence of a non-trivial bluer accretion disk component in the optical emission which also affects the slope of the OIR spectrum. This may explain the overlap of the color-magnitude behavior of intervals 1 and 2 in Figure \ref{colormag}.

The X-ray emission from FSRQs is often dominated by the SSC process \citep{muk99,har01,sik01,cha08} which means that changes in magnetic field should lead to X-ray variability at a level similar to those changes. This is roughly consistent with the relative variability in optical and X-ray flux seen here. We note that the optical synchrotron emission and the SSC X-rays are produced by different parts of the electron distribution. Assuming a B field of 1 Gauss, synchrotron emission at optical wavelengths is generated by electrons of Lorentz factor $\gamma$ $\sim$ $10^4$ while SSC X-rays are produced by $\gamma$ $\sim$ $10^2$ assuming average seed photons to be at the IR wavelength range where the synchrotron peak occurs for most FSRQs. The difference in the OIR and X-ray variability, if any, in 0208-512 during interval 2 can be attributed to this.

In another possible scenario, the large difference in the GeV/OIR ratio between intervals 1 and 3 vs interval 2 may be caused by the location of the outbursts. The acceleration and collimation of magnetohydrodynamically launched jets occur over an extended region \citep{mei00,vla04,mck09}, possibly as high as $\sim$$10^4$ times the Schwarszchild radius. Hence, an outburst taking place closer to the black hole may be associated with a smaller bulk Lorentz factor ($\Gamma$). In contrast the corresponding magnetic field and particle density will be larger due to its compactness. It has been shown that the EC/synchrotron ratio can be dramatically different depending on the location where a given outburst takes place, with EC becoming relatively more dominant for outbursts occurring farther down the jet \citep{kat07}. The apparent seed photon field seen by the emitting electrons is boosted by a factor of $\Gamma^2$ causing a significant dependence of the EC/synchrotron ratio on $\Gamma$. If the outburst during interval 2 is generated closer to the BH where $\Gamma$ is smaller while the outbursts in intervals 1 and 3 are produced farther down the jet where $\Gamma$ is larger, the optical--near IR emission will be significantly more dominating than that in the GeV bands during interval 2 than the other two. A factor of two decrease in $\Gamma$ coupled with a factor of two increase in B can produce the change of EC/synchrotron ratio we see from interval 1 to 2.

In this alternative scenario, the SSC X-rays increase when the EC component is weaker because a  larger fraction of the total dissipation occurs through SSC process. However, the contribution of the EC component at the X-ray energies decrease at the same time causing the observed X-ray flux to be similar or even lower depending on specific parameters of the source \citep{kat07}, as observed here.

This is one of the very few cases where one of two specific physical scenarios can be identified for certain observed events in a blazar jet. These events can be ideal candidates for detailed modeling of the time variable SED to provide stronger constrains on the relevant physical parameters. This will be addressed in a future paper.

\section{Acknowledgments}
RC received support from \textit{Fermi} GI grant NNX09AR92G. SMARTS observations of LAT-monitored blazars are supported by Fermi GI grant NNX10A043G and NNX12AP15G. CDB, MMB and the SMARTS 1.3m observing queue also received support from NSF grant AST-0707627. GF is supported by \textit{Fermi} GI grant NNX10A042G. JI is supported by the NASA Harriet Jenkins Fellowship program and NSF Graduate Research Fellowship under Grant No DGE-0644492. This work made use of data supplied by the UK \textit{Swift} Science Data Center at the University of Leicester.

\end{document}